\documentstyle[preprint,revtex,eqsecnum]{aps}
\begin{document}
\begin{title}
The Quantum Hall Effect in Drag: Inter-layer \\
Friction in Strong Magnetic Fields
\end{title}
\author{
Efrat Shimshoni$^{(a,b)}$
and
S.\ L.\ Sondhi$^{(b)}$
}
\begin{instit}
$^{(a)}$Beckman Institute,
405 North Mathews Avenue,\\
$^{(b)}$Department of Physics,
1110 West Green Street,\\
University of Illinois at Urbana-Champaign,
Urbana, Illinois 61801-3080, USA
\end{instit}

\begin{abstract}

We study the Coulomb drag between two spatially separated electron
systems in a strong magnetic field, one of which exhibits the
quantum Hall effect.  At a fixed temperature, the drag
mimics the behavior of $\sigma_{xx}$ in the quantum Hall system, in that it is
sharply peaked near the transitions between  neighboring
plateaux.
We assess the impact of critical fluctuations near the transitions,
and find that the low temperature behavior of the drag measures an
exponent $\eta$ that characterizes anomalous low frequency
dissipation; the latter is believed to be present following the work of
Chalker.

\end{abstract}

\pacs{PACS numbers:
	73.50.Dn 
        72.20.My 
        73.20.Dx 
        71.55.Jv 
        71.30.+h 
}
\newpage
\narrowtext
Coulomb interactions between spatially separated systems of
charge carriers lead to a variety of novel phenomena
\cite{REF:DLQH}.
In particular, density fluctuations in one of the
systems exert a frictional force on the other, and consequently
a current flowing in one induces a drag current in the other. This
effect was first predicted by Pogrebinskii and by Price
\cite{REF:Pog}.
A number of recent experiments have successfully observed the
Coulomb drag in heterostructures involving a two-dimensional electron system
(2DES), separated from another electron \cite{REF:SolEx,REF:Gramila}
or hole \cite{REF:Sivan} gas by an
insulating barrier sufficiently thick to prevent tunneling. Solomon et al.
\cite{REF:SolEx} observed the drag in a three-dimensional electron gas (3DEG)
in a GaAs gate electrode, while Gramila et al. \cite{REF:Gramila} and Sivan
et al. \cite{REF:Sivan} studied double layer structures at lower temperatures.
This experimental progress has been accompanied by a number of
theoretical studies
\cite{REF:Laikh,REF:Jauho,REF:Tso,REF:ZM}.

In this letter we study the Coulomb drag between two electron
systems, one of which is two dimensional and exhibits
the quantum Hall effect (QHE) \cite{REF:QHE}, and the other,
christened the test system, could (in principle) be in any state feasible
in a magnetic field.  (We shall concentrate on the cases where it is
either an identical layer or is a semi-infinite 3DEG.)
As we will illustrate, the Coulomb drag in this situation is a novel probe
of the internal dynamics of the 2DES in the QH regime.
The geometry we consider is depicted in Figure 1; the test system is situated
above the QH layer, across an insulating barrier that suppresses
tunneling but permits Coulomb scattering \cite{REF:SolEx,REF:Gramila}.
The quantity of interest is the trans-resistivity $\rho_t$
defined as
\FL
\begin{equation}
\rho_t={\cal E}_u/ j_d\, ,
\label{EQ:Rt}
\end{equation}
where ${\cal E}_u$ is the parallel electric field induced in U (which is in
an open circuit) in response to a current density $j_d$ established in D.
We find that the
dependence of $\rho_t$ on the filling factor $\nu$ at a fixed temperature
is similar to that of the dissipative conductance $\sigma_{xx}$
in that it peaks at the transition between QHE plateaux and is greatly
suppressed deep in the QH phases;
this pattern arises independently of whether the test system exhibits the QHE.
This is quite
intuitive: as with everyday friction, the magnitude of the drag is
directly related to inhomogeneities in the systems---in our problem,
long lived density fluctuations. Deep in the QH phases these are
suppressed (``incompressibility'') but they grow as the boundary of the phase
is approached.

An important consequence of this physics is that the
$T$-dependence of $\rho_t$ is sensitive to the low frequency dynamics of
the density fluctuations and is different at different filling
factors. In particular, for macroscopic systems
Zheng and MacDonald \cite{REF:ZM} have shown
that $\rho_t$ involves a convolution of the dissipative density-density
response functions ${\rm Im}\chi(q,\omega)$ of the individual layers
(Eq.~(\ref{EQ:RtUD})). As a result,
$\rho_t$ at low $T$ is sensitive to ${\rm Im}\chi$ for $\omega <q$,
{\em in contrast} to dc transport measurements in a single layer
that probe $q<\omega$.
Most strikingly, for two identical layers we find that while
$\rho_t \sim O(T^2)$ in the
QH phases, at the critical points $\rho_t \sim T^{2-\eta}$, where $\eta$
is an exponent characteristic of anomalous low-frequency dissipation, first
discussed, in the context of
the anomalous diffusion at the non-interacting critical point for the IQHE,
by Chalker \cite{REF:chalk1}.
Such a direct measurement of this exponent, which is not
experimentally accessible by other means, could shed
considerable light on the nature of the critical fixed points and the issue
of the universality of the various transitions in the QH regime.

The paper is organized as follows.
First, we discuss a model of a clean system, applicable to quantum wires,
where our scenario is realized transparently by means of
Fermi's golden rule.
Next, we extend our considerations to disordered systems,
particularly focusing on the critical region where the discussion relies on
known or conjectured properties of ${\rm Im}\chi$. In
most of the following we shall discuss the IQHE, even though the physics is
evidently more general; the extension to the
FQHE is discussed at the end of the paper.

\noindent
{\bf Clean System:} We assume that layer D is free of impurities and
is subject to a parabolic confining potential,
${1\over 2}m\omega_0^2 y^2$. The exactly calculable eigenstates \cite{REF:MEG}
are labeled by the
wave-vector in the $x$-direction $k_d$, and the Landau level (LL) index
$n_d$; the corresponding energy levels are denoted $E^{(d)}_{n_d,k_d}$.
The electron system in layer U is assumed to be a 3DEG; its eigenstates are
labeled by $n_u$, $k_u$ and $k_z$, and the energy
levels are denoted $E^{(u)}_{n_u,k_z}$.
We calculate $\rho_t$ within the Boltzmann transport approach. The current
density $j_d$ flowing in D along the $x$-direction is
\FL
\begin{equation}
j_d={{e\hbar}\over{\pi L_y m^\ast}}\sum_{n_d}
\int_{-\infty}^\infty dk_d\, g_{n_dk_d}k_d\; ,
\label{EQ:jd}
\end{equation}
where $g_{n_dk_d}\equiv-\hbar v_D(\partial f^0_{n_dk_d}/\partial
E^{(d)}_{n_d,k_d})k_d$ is the deviation of the electron distribution from
the equilibrium Fermi distribution $f^0_{n_dk_d}$, $v_D$ is the drift
velocity, and $m^\ast\equiv m(\omega_c/\omega_0)^2$ in terms of the band
effective mass $m$ and the cyclotron frequency $\omega_c$.
The electric field ${\cal E}_u$ induced in U is related to
the total momentum transfer into this region by
\FL
\begin{equation}
{\cal E}_u={{\hbar L_x L_z^\ast}\over{e\pi^2 N_u }}
\sum_{n_u}\int_0^\infty dk_z
\int_{-\infty}^\infty dk_u\, k_u\tau^{-1}(k_u, k_z)\; ,
\label{EQ:Eu}
\end{equation}
where the relaxation rate $\tau^{-1}(k_u, k_z)$ is, to linear order in
$g_{n_dk_d}$,
\FL
\begin{eqnarray}
&&
\tau^{-1}(k_u, k_z)={2\pi\over\hbar}{L_z^\ast L_x^3\over 4\pi^4}
\sum_{n_u^\prime,n_d^\prime,n_d}\int_0^\infty dk_z^\prime
\int_{-\infty}^\infty dk_u^\prime\int_{-\infty}^\infty dk_d^\prime
\int_{-\infty}^\infty dk_d |V_{ud,u^\prime d^\prime}|^2
\nonumber
\\
&&
\quad\times\{P(u,d;u^\prime,d^\prime)-P(u^\prime,d^\prime;u,d)\}
\delta(E^{(d)}_{n_d,k_d}+E^{(u)}_{n_u,k_z}-E^{(d)}_{n_d^\prime,k_d^\prime}
-E^{(u)}_{n_u^\prime,k_z^\prime})\; ,
\nonumber
\\
&&
P(u,d;u^\prime,d^\prime)\equiv f^0_{n_uk_z}(1-f^0_{n_u^\prime,k_z^\prime})
\{g_{n_dk_d}(1-f^0_{n_d^\prime,k_d^\prime})-
g_{n_d^\prime,k_d^\prime}f^0_{n_dk_d}\}\; ;
\label{EQ:Rr}
\end{eqnarray}
here $N_u$ is the number of electrons in U, and
$V_{ud,u^\prime d^\prime}$ is the matrix element of the screened
Coulomb interaction across the barrier \cite{REF:Lzeff}.

We now analyze the variation of ${\cal E}_u$ near the transition
between successive plateaux, {\em i.e.} when either the density or $\bf B$
is varied so that the chemical potential $\mu$
crosses the bottom of a LL. The interesting region is when
$\mu$ lies just above the bottom of the highest occupied LL, where
we can ignore scatterings between different Landau levels.
For our qualitative purposes it is sufficient to consider the case where the
highest occupied levels are the lowest LL in both D and U.
The crucial step
is to distinguish two different regimes in the position of $\mu$:
$(a)$ $k_BT\ll E_F^{(d)}$ and $(b)$ $k_BT\gg E_F^{(d)}$, where
$E_F^{(d)}\equiv(\mu-E^{(d)}_{00})$. A bias between the layers is chosen
so that in both regimes
$E_F^{(u)}\equiv(\mu-E^{(u)}_{00})\gg k_BT$. Correspondingly, we define
$k_F^{(d)}\equiv\sqrt{2m^\ast E_F^{(d)}}/\hbar$ and
$k_F^{(u)}\equiv\sqrt{2m E_F^{(u)}}/\hbar$.

In regime $(a)$, the electron gas in the layer is characterized by a sharp,
two points Fermi ``surface'' $k_d=\pm k_F^{(d)}$.
At $T=0$,
$\tau^{-1}(k_u, k_z)$ vanishes (cf. Eq.~(\ref{EQ:Rr})); at small
$T$, it is dominated by forward ($k_d\approx k_d^\prime$)
and backward ($k_d-k_d^\prime\approx\pm 2k_F^{(d)}$) scattering processes.
The former are negligible, and the latter are
$\sim\exp[-(2k_F^{(d)}\ell)^2]\exp[-4k_F^{(d)}d]$ \cite{REF:qsln}.
For $E_F^{(d)}>max\{\hbar^2/2m^\ast(2\ell)^2,\hbar^2/2m^\ast(4d)^2\}$,
$\rho_t$  is severely suppressed. As $E_F^{(d)}$ is reduced below
$min\{\hbar^2/2m^\ast(2\ell)^2,\hbar^2/2m^\ast(4d)^2\}$,
$\rho_t$ becomes more appreciable and approaches its maximal
value as $E_F^{(d)}$ is further reduced, crossing over to regime $(b)$.

In regime $(b)$, where $\rho_t$ is peaked,
we assume the hierarchy of energy scales
$E_F^{(d)}\ll k_BT\ll E_F^{(u)}$. The 2DES becomes an effectively
classical gas, where $k_BT$ replaces
$E_F^{(d)}$ as the typical electron energy. Consequently, one should
account for the $T$-dependence of $q_s$, the screening wave-vector in  D,
as well as the effective width of this layer.
The latter is given, roughly, by the effective real-space extent
of the electronic states with energy lower than
$E^{(d)}_{00}+k_BT$. We thus obtain the peak trans-resistivity
\FL
\begin{equation}
L_z^\ast\rho_t^{(0)}\approx{3\pi^5\over 16}{\hbar\over e^2}\Bigg({a_B\over d}
\Bigg)^2 {\epsilon^2(k_F^{(u)})^5\ell^8 (k_BT)^2\over e^4}\; ,
\label{EQ:Peak}
\end{equation}
where $a_B\equiv \hbar^2/e^2m$,
$\ell$ is the magnetic length and $\epsilon$ the background dielectric
constant. For $m\approx 0.07m_e$ and $\epsilon\sim 10$ (appropriate to GaAs),
$d=200\,{\AA}$, $\ell=100\,{\AA}$, $k_F^{(u)}\sim 1/\ell$ and $T=1\,K$, we get
$L_z^\ast\rho_t^{(0)}\sim 10^{-8}\,\Omega {\rm cm}$.

\noindent
{\bf Systems with disorder:}  We now turn to realistic systems where, in
contrast to the clean example, there are
always states at the Fermi level in the bulk. We
distinguish two regions: the critical region, where
electron delocalization dominates, and the non-critical
region.

In the latter, the scattering between the layers and hence the drag may be
assisted by hopping among localized states, in addition to the edge-edge
processes considered earlier.
As with the treatment of $\sigma_{xx}$ in the framework of variable range
hopping \cite{REF:ps}, these contribute
$\rho_t \sim e^{-(T_0/T)^\alpha}$,
where the exponent $\alpha=1/3,1/2$ for the cases of Mott hopping and Coulomb
gap physics respectively. We find that for reasonable system sizes this
form holds, except at the very
lowest temperatures where the range of the hopping exceeds the sample size,
and the edge-edge scattering dominates; for a sample of size $L \sim 10
\mu m$, we find that the crossover temperature is at most $10^{-4}K$.

\noindent
{\bf Critical Region:} In this region the system is in the vicinity of
a $T=0$ critical
point with a finite non-zero $\sigma_{xx}$. At the latter,
$\rho_t$ will be dominated by the extended states that exist
in the bulk of the sample.
This bulk contribution is conveniently evaluated using the result of
Zheng and MacDonald \cite{REF:ZM}, who showed that to leading non-trivial
order in the interaction $V$  between two 2DESs,
\FL
\begin{equation}
\rho_t={\beta\hbar^2\over \pi n^{(u)}
n^{(d)}e^2}\int{d^2 q \over(2\pi)^2}q^2|V(q)|^2\int_0^\infty d\omega
{{\rm Im}\chi_d(q,\omega)
{\rm Im}\chi_u(q,\omega)\over 4{\rm sinh}^2(\beta\hbar\omega/2)}\; ,
\label{EQ:RtUD}
\end{equation}
where ${\rm Im}\chi_{d(u)}(q,\omega)$ are the dissipative density-density
response functions of layer D (U)\cite{fn2}.
Two features of Eq.~(\ref{EQ:RtUD}) are worth noting. First,
the leading low $T$ behavior of $\rho_t$ can be obtained by
approximating the response functions by their $T=0$ forms. Second, as
$T \rightarrow 0$, the $T$ dependence of the denominator forces the relevant
range of $\omega$ to vanish, and hence the behavior of $\rho_t$ becomes
sensitive to the form of ${\rm Im}\chi_{d(u)}(q,\omega)$ for $\omega <q$.

In applying Eq.~(\ref{EQ:RtUD}) to the critical region we are faced with the
problem that a proper theory of the latter
is not available at present.
Therefore we take a dual approach. First, for the well studied
non-interacting problem,
which appears to be relevant to the
experiments thus far \cite{REF:scaling}, we show that $\rho_t$ vanishes as a
universal
power of $T$. Next, we show that this behavior is generic; at {\em any}
(interacting) critical point  $\rho_t$ measures an exponent characteristic of
the corresponding universality class. Hence, measuring $\rho_t$
would test the non-interacting theory, as well as serve
as a model-independent probe of the universality classes among the QH phase
transitions.

For non-interacting electrons Chalker and Daniell \cite{REF:chalk-dan}
have shown that the low frequency
response at criticality has the form:
\FL
\begin{equation}
{\rm Im}\chi_d(q,\omega)={(dn^{(d)}/d\mu)\omega {\cal D}(q,\omega)
q^2\over [\omega^2+({\cal D}(q,\omega)q^2)^2]}\; ,
\quad {\cal D}(q,\omega)= D \Bigg({{\cal C}\omega\over q^2}\Bigg)^{\eta/2}.
\label{EQ:anchi}
\end{equation}
(For ${\cal C}\omega>q^2$,
${\cal D}(q,\omega) \approx D=0.087/\hbar(dn^{(d)}/d\mu)$;
${\cal C}\approx 60\hbar(dn^{(d)}/d\mu)$ where $dn^{(d)}/d\mu$ is
the density of states at the band center, and
the universal exponent $\eta \approx 0.38$). On substituting
this in Eq.~(\ref{EQ:RtUD}) we obtain, for two identical layers
\FL
\begin{equation}
\rho_t\approx 3.1\times 10^{-4} {h\over e^2}
\Bigg({k_B T\over\hbar Dn^{(d)}}\Bigg)^2
\Bigg({\epsilon\over e^2d(dn^{(d)}/d\mu)}\Bigg)^2
\Bigg({{\cal C}d^2 k_BT\over \hbar}\Bigg)^{-\eta}\; .
\label{EQ:RtT}
\end{equation}
Hence, in this scenario, $\rho_t$ in the critical region is {\em
parametrically} enhanced at low $T$
over non-critical filling factors, and its $T$ dependence directly measures
$\eta$ (with a non-universal prefactor).
As noted by Chalker \cite{REF:chalk2}, the non-zero value of
$\eta$ reflects the presence of large amplitude fluctuations in the critical
eigenstates. Also, as shown in \cite{REF:huck},
it leads to anomalously slow relaxation of the local density fluctuations.
Hence, the enhancement of $\rho_t$ is entirely in accord with our
earlier discussion of the physics. For a sample of mobility
$\sim 10^6\,{\rm cm^2/Vs}$, $n^{(d)}\sim10^{11}\,{\rm cm}^{-2}$,
$d=200\,{\AA}$ and $B=10T$,
we estimate $\rho_t \approx 10\,{\rm m}\Omega $ at $T \approx 0.1 K$.
As with the scaling of $\sigma_{xx}$, the behavior of $\rho_t$ becomes
non-critical at a crossover temperature $T^\ast \sim 1/\xi$ \cite{REF:scaling}.

For completeness we note that for the
case where U is three dimensional, $\rho_t\sim T^{2-\eta/2}$.
Thus, $\rho_t$ vanishes faster with $T$ than in Eq.~(\ref{EQ:RtT}), and
we find that to observe an appreciable
sensitivity to $\eta$,  $T$ has to be
reduced to $\sim 1\ mK$.

We now consider the situation at an arbitrary interacting critical point,
with some unknown values of the localization length exponent $\nu_\xi$ and
the dynamic scaling
exponent $z$.
Fortunately, ${\rm Im}\chi$ is the correlator of a conserved density
and
hence by standard renormalization group arguments it has
the critical scaling form:
\begin{equation}
{\rm Im}\chi(q,\omega) = q^{d-z} f(\omega/q^z).
\end{equation}
${\rm Im}\chi$ is odd in $\omega$, and away from criticality we expect
${\rm Im}\chi \sim \omega f(q)$ at small $\omega$ for fixed $q$. Assuming that
the critical fluctuations modify this linear dependence we parametrize the
deviation by postulating  $f(x) \sim x^{1 - \eta/2}$ for $x \ll 1$ \cite{fn1}.
For our problem ${\rm Im}\chi \sim \omega^{1 - \eta/2} q^{2-2z +
\eta z/2}$
for $\omega \ll  q $; this agrees with the non-interacting form with $z=2$.
Substituting this in Eq.~(\ref{EQ:RtUD}) gives
$\rho_t \sim T^{2- \eta}$ at low $T$. Hence an
enhanced drag at criticality
would still measure an exponent characteristic of anomalous low frequency
dissipation \cite{REF:extend}.

Finally, we comment on the FQHE. The
qualitative variation of $\rho_t$ with filling factor at fixed $T$ derives
from variations in the compressibility, which are
similar in the IQHE and the FQHE. The behavior of the clean system
should be similar, even though one has
Luttinger liquids at the edges, for their small $q$ density-density response
is the same as that of the integer Fermi liquids. For disordered
samples, some form of variable range hopping should hold deep in the
plateaux, where the fractional statistics of the quasiparticles
is likely unimportant. As for the critical region, even less is known than
for the IQHE; hence, as already emphasized, we expect measurements of
$\rho_t$, and hence of $\eta$, to be particularly illuminating.

\acknowledgements

We are very grateful to A.H. MacDonald for much useful advice and
for informing us of his work.
We thank G.\ Baym, J.\ Eisenstein,
J.\ Engelbrecht, E.\ Fradkin,
S.\ Girvin, A.\ Jauho, B.\ Laikhtman,
A.\ Leggett and B.\ Shklovskii for
helpful discussions; and A.H. MacDonald and S. Kivelson for comments on the
manuscript. This work was supported in part by the
Beckman Foundation (ES), and NSF grants No. DMR 91--22385,
DMR 91-57018 and the Aspen Center for Physics (SLS).

\figure{\label{fig:hetero}}
Schematics of the heterostructure. The shaded area is an insulator, and
{\bf B} is the magnetic field.

\begin{references}
\bibitem{REF:Lzeff}
$L_z^\ast$ is the effective thickness of layer U; for
$L_z<\lambda_{sc}$ (the screening length in U), $L_z^\ast\approx L_z$, and for
$L_z>\lambda_{sc}$, $L_z^\ast\approx\lambda_{sc}$.
\bibitem{REF:qsln}
We note that near $q=2k_F^{(d)}$ and at finite $T$ there is an algebraic
enhancement of the interaction due to the reduction of the screening
wave-vector $q_s(q=2k_F^{(d)})\sim\exp[-(k_F^{(d)}\ell)^2]
\ln[{8E_F^{(d)}/ k_BT}]$. However, this minor enhancement is overwhelmed
by the exponentials.
\end{references}
\end{document}